\documentclass[12pt, draftclsnofoot, onecolumn]{IEEEtran}
\usepackage{color}
\usepackage{cite}
\usepackage{amsmath}
\usepackage{algorithmic}
\usepackage{algorithm}
\usepackage{array,amsfonts}
\usepackage{setspace}
\usepackage{eqparbox}
\usepackage{amssymb,pst-poly}
\usepackage{bm}
\usepackage{graphicx}
\usepackage{subfigure}
\usepackage{stfloats}
\usepackage{caption}

 \usepackage{epstopdf}

\usepackage{mathrsfs}
\makeatletter \@addtoreset{equation}{section} \makeatother

\begin{document}

\title{Fast Authentication and Progressive Authorization in Large-Scale IoT: How to Leverage AI for Security Enhancement?}

\author{He~Fang,~\IEEEmembership{Student Member, IEEE}, Angie~Qi,~and~Xianbin~Wang,~\IEEEmembership{Fellow, IEEE}

\thanks{H. Fang and X. Wang are with the Department of Electrical and Computer Engineering, The University of Western Ontario, London, ON N6A 5B9, Canada. Email: hfang42@uwo.ca, xianbin.wang@uwo.ca.}

\thanks{A. Qi is with the Faculty of Science, The University of British Columbia, Vancouver, BC V6T 1Z1, Canada. Email: angieqi.5848@gmail.com.}

\thanks{This
work was supported  by the NSERC Discovery Program under project number RGPIN-2018-06254. (Corresponding
author: Xianbin Wang).}
}

\maketitle

\begin{abstract}
Security provisioning has become the most important design consideration for large-scale Internet of Things (IoT) systems due to their critical roles to support diverse vertical applications by connecting heterogenous devices, machines and industry processes. Conventional authentication and authorization schemes are insufficient in dealing the emerging IoT security challenges due to their reliance on both static digital mechanisms and computational complexity for improving security level. Furthermore, the isolated security designs for different layers and link segments while ignoring the overall protection lead to cascaded security risks as well as growing communication latency and overhead. In this article, we envision new artificial intelligence (AI) enabled security provisioning approaches to overcome these issues while achieving fast authentication and progressive authorization. To be more specific, a lightweight intelligent authentication approach is developed  by exploring machine learning at the gateway to identify the access time slots or frequencies of resource-constraint devices. Then we propose a holistic authentication and authorization approach, where online machine learning and trust management are adopted for analyzing the complex dynamic environment and achieving adaptive access control. These new AI enabled approaches establish the connections between transceivers quickly and enhance security progressively, so that communication latency can be reduced and security risks are well-controlled in large-scale IoT. Finally, we outline several areas for AI-enabled security provisioning for future researches.
\end{abstract}

\IEEEpeerreviewmaketitle

\section{INTRODUCTION}

The advent of large-scale Internet-of-Things (IoT) systems and their ongoing convergence with diverse industry applications signify the imminent next wave of the ubiquitously connected society \cite{06}.
However, the \textit{complexity} of large-scale IoT systems as well as dramatically increased use of \textit{intelligent machines and devices} within industry processes
bring many security vulnerabilities and new design challenges.
As shown in Fig. 1, potential security risks and attacks could lead to catastrophic consequences and cause avalanche-like damages in large-scale IoT networks. This is mainly due to the critical roles of IoT to support a wide variety of vertical applications by connecting tremendous \textit{heterogenous} devices, machines and industry processes, as well as cascaded reaction from the enormous parallel interconnection contained in IoT.
Moreover, the widely used \textit{resource-constrained devices}, e.g. sensors, can be compromised easily, thus resulting in  widely distributed threats to the IoT network through data injection, spoofing, eavesdropping, and so on.
With the cascaded effect, these security threats  could also lead to the failure of a whole large-scale IoT system, especially for those applications relying on tight collaboration among diverse entities.

\begin{figure}[htbp]
\centering
\includegraphics[width=16cm,height=8.5cm]{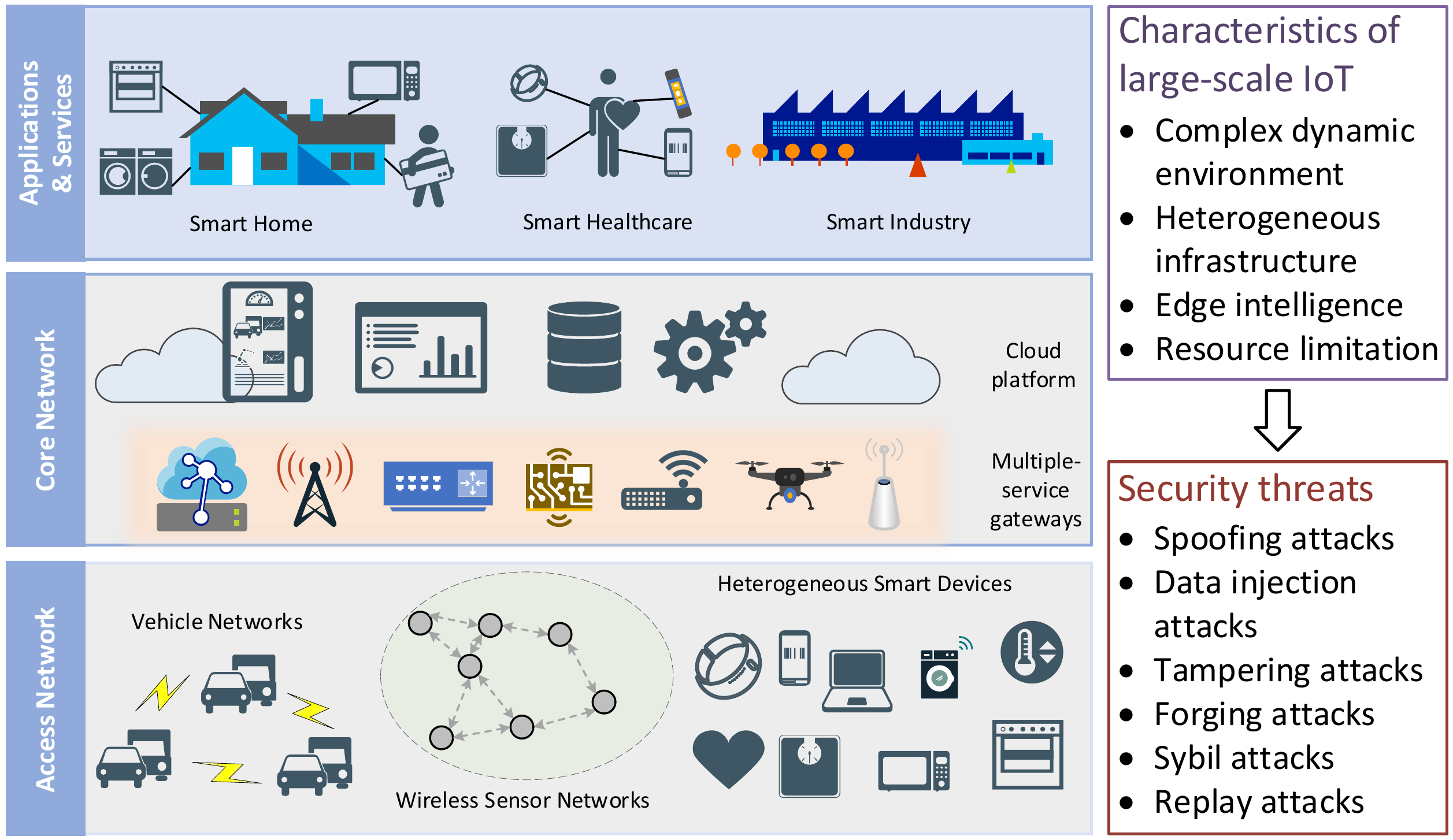}
\caption{Network architecture of large-scale IoT and  typical security threats.}
\end{figure}

Authentication and authorization have been considered as  key security mechanisms and critical designs for large-scale  IoT networks, since adversaries need the access to IoT systems to launch attacks  \cite{06,09}.
These mechanisms secure the legitimate communications through confirming the identities of all devices and their right access to the authorized resources, data and services. However, the specific characteristics of large-scale IoT networks bring new design challenges for security provisioning. Explicitly, those IoT devices suffering from  resource limitation could
not support the security methods  requiring high communication and computation overhead, and tremendous devices contained in a large-scale IoT system claim low-delay transmissions to guarantee their communication performance.
Hence, to defend against security threats in large-scale IoT networks, this article focuses on studying the challenges for conventional authentication and authorization schemes, and envisioning new approaches for security enhancement.

\subsection{Challenges for Conventional  Authentication and Authorization in Large-Scale IoT}
The conventional authentication and authorization methods, including key-based cryptography techniques and physical layer key generation techniques,  may suffer from their high complexity and long latency, and may be ineffective to adapt to the complex dynamic environment, especially in large-scale IoT networks. More importantly, their generated keys may be leaked  in the security management procedures, e.g. key distribution. Their challenges  in supporting  large-scale IoT applications are summarized as follows.\\
\textbf{Long security induced latency in large-scale IoT.}
The conventional cryptography techniques require increased overhead and lengthy process for increased level  of security, thus leading to high communication and computation overhead, more importantly,  long communication latency.
These are intolerable for the large-scale IoT network having significantly increasing number of intelligent machines and resource-constrained devices with concurrent communications. Moreover, the conventional statistical methods for authentication also require enough time and  computational resources for obtaining the statistical properties, thus leading to limited capability in detecting attacks instantaneously. As a result, new fast and lightweight authentication scheme is necessary for large-scale IoT applications. \\
\textbf{Ineffective adaptation to complex dynamic IoT environment.} Conventional security solutions may also
suffer from cascading risks in complex dynamic IoT scenarios due to their reliance on static binary authentication/authorization mechanisms. Such mechanisms are difficult in learning from and adapting to the complex dynamic IoT environment encountered, thus resulting in the failure of continuous protections for legitimate communications and decrease of security performance in dealing with varying security risks. Hence, new concept of continuous progressive authorization is extremely beneficial for holistic system optimization in complex dynamic environments.\\
\textbf{Potential key leakage in the security management procedures.} Conventional cryptographic techniques also require appropriate key management procedures to generate, distribute, refresh and revoke digital security
keys, leading to the potential key leakage. Furthermore, in the physical key generation techniques \cite{02}, the key transmission is  needed for information reconciliation. These all result in tremendous loopholes for adversaries and widely security threats in large-scale IoT systems. In achieving the privacy protection in security management, new mechanism should be designed for intelligent security provisioning.

In a nutshell, the security enhancement  is of paramount importance for large-scale IoT networks, especially in the coming of information age requiring ``intelligence".

\subsection{Artificial Intelligence for Security Enhancement in Large-Scale IoT}
The artificial intelligence (AI) \cite{04} is leveraged  in this article to overcome the above challenges and to enhance security in large-scale IoT networks through learning, reasoning, and self-correction.
Explicitly, AI techniques could be exploited at gateways and routers, and the abundant information contained in large-scale IoT systems could be utilized for learning and reasoning, thus achieving the lightweight  and fast authentication. Moreover, through self-correction by AI techniques, the progressive authorization may be accomplished to adapt to complex dynamic IoT environments.
To be more specific, AI may contribute to the security enhancement in large-scale IoT  networks due to following reasons.
\begin{itemize}
\item \textbf{Security management may be accelerated by AI based on multi-domain information in large-scale IoT.} In a large-scale IoT system, the gateways and routers may undertake the AI management, such as data collection, training and testing,
    thus the communication and computation overhead could be reduced at low-power devices. It also contains abundant multi-domain information of communication channels, devices, environments, network connections, and application softwares. Such information could be intelligently utilized for security provisioning as well as for contributing to the learning and reasoning  by AI techniques.  More importantly, AI techniques may utilize the historical  information to facilitate the security management. Hence, the continuous authentication and authorization of devices could be accelerated as well as the security induced latency could be reduced by Al based on multi-domain information.
\item \textbf{AI provides real-time learning under limited statistical properties and unpredictable dynamics.} In the practical communication environment, it is more and more difficult to develop accurate statistical models for  security enhancement. This is mainly because of the ubiquitous uncertainties and unpredictable dynamics as well as the limited computational resources and  time for obtaining the precise statistical properties. Those AI techniques providing online learning under limited statistical properties and unpredictable dynamics could conduce to the real-time detection of attacks and continuous protections for legitimate communications. Furthermore, different from the statistical hypothesis testing \cite{12}, which is a method of statistical inference, the machine learning algorithms allow software applications to become more accurate in predicting outcomes  based on the training data without being explicitly programmed.
    \item \textbf{Privacy protection in the security management may be achieved with the help of AI techniques.}  To achieve security enhancement, the authentication and authorization mechanisms should be protected from the privacy leakage. With the help of AI techniques, intelligent security provisioning may be designed through utilizing the channel reciprocity \cite{02} as well as  the specific communication link-related, device-related, and biometric features without the transmission of  private information. To be more specific, AI techniques may help in amplifying the channel reciprocity, so that the highly similar private information can be acquired on transmitter and receiver sides. Moreover, those machine learning algorithms may track the specific features for the continuous authentication without the transmission of private information.
\end{itemize}
Therefore, through leveraging AI, new intelligent security provisioning schemes could be developed in overcoming the specific challenges in large-scale IoT networks.

The rest of this article goes as follows. The existing AI-based authentication schemes are reviewed firstly. Then we envision a lightweight authentication scheme as well as a holistic authentication and authorization scheme based on AI techniques for large-scale IoT networks. These approaches provide paradigms for  security enhancement by leveraging AI to achieve fast authentication and progressive authorization.  Finally,  this article is concluded and future research perspectives are outlined.

\section{STATE OF THE ART ON AI-BASED AUTHENTICATION}
With the rapid developments of smart applications and automatic techniques in the new era,  IoT security and AI have attracted a lot of attentions from both academic and industrial communities. Although IoT security has been studied in the literature \cite{06,09}, new perspectives of AI-based solutions will be beneficial for security enhancement, especially in complex dynamic  large-scale IoT networks.
A kernel machine learning-based physical layer authentication scheme is proposed in \cite{01} to defend against spoofing attacks through tracking the communication link and hardware-related features in time-varying environments. \cite{12} develops a physical-layer authentication scheme based on the extreme learning machine to improve the spoofing detection accuracy.
To detect cyber-attacks, \cite{07} designs  a watermarking algorithm based on a deep learning long short-term memory structure for dynamic authentication, which enables the IoT devices to extract a set of stochastic features from their generated signal and to dynamically watermark these features into the signal.
The authors in \cite{08} proposed   a physical unclonable function (PUF) based on current
mirror array circuits that reuses the circuit implementation of  the extreme learning machine for authentication.
The deep learning technique is exploited in \cite{10} to recognize the behavior features of malicious data injection attacks relying on the study of historical measurement data and captured
features.
In \cite{13},  a learning-based model is designed for the central server to extract mobility features and distinguish Sybil attackers from
benign vehicles through analyzing their mobility behaviors.

Although the above schemes achieve security enhancement by exploring AI techniques, they are insufficient in overcoming the specific challenges presented above.
In addition, the extra computation and communication overhead as well as long time-latency may be increased due to the lengthy processes of training, testing and attack detection by AI techniques, thus moving far away from the fast and lightweight authentication.
More importantly, most of these schemes are also binary in nature, which means a device either pass the authentication and access all the authorized services/resources, or fail the authentication. This is incapable of providing the progressive and robust protections for legitimate communications.  Hence, we focus our attention on envisioning new AI-aided security approaches  to overcome the specific challenges as well as to achieve the fast authentication and progressive authorization in large-scale IoT networks. These AI-aided
solutions will provide a new insight of leveraging AI for  security enhancement.


\section{AI-ENABLED FAST AUTHENTICATION AND PROGRESSIVE AUTHORIZATION}
In this article, we envision new AI-enabled paradigms in large-scale IoT for introducing intelligence to security management and for achieving fast authentication and progressive authorization, including a lightweight authentication scheme as well as  a holistic authentication and authorization scheme.
In these schemes, AI contributes to the security enhancement
not only by seeking and utilizing valuable  information in large-scale IoT systems, but also through
providing self-adaptation for authentication and authorization.

\subsection{Lightweight Authentication Exploiting Support Vector Machine (SVM)}

We first envision a lightweight authentication scheme for identifying multiple sensors according to their pseudo-random accesses  in  time-domain or in frequency-domain. To be specific, only when the access time slots or access frequencies of a sensor are identical to its unique pseudo-random binary sequence (PRBS), it will be authenticated as legitimate device by the gateway.  The seed for generation of a PRBS between each sensor and the  gateway can be obtained by utilizing their  unique features, as exemplified by physical layer attributes \cite{01}, as well as by exploring the support vector machine (SVM) \cite{14} to nonlinearly classify the measurements of features.  Hence, this scheme provides fast authentication by identifying the access time slots or frequencies directly and
progressive protections for the legitimate communications without requiring complex computation and high communication abundant at sensors, demonstrating a lightweight authentication scheme.

\begin{figure}[htbp]
\centering
\includegraphics[width=12cm,height=10cm]{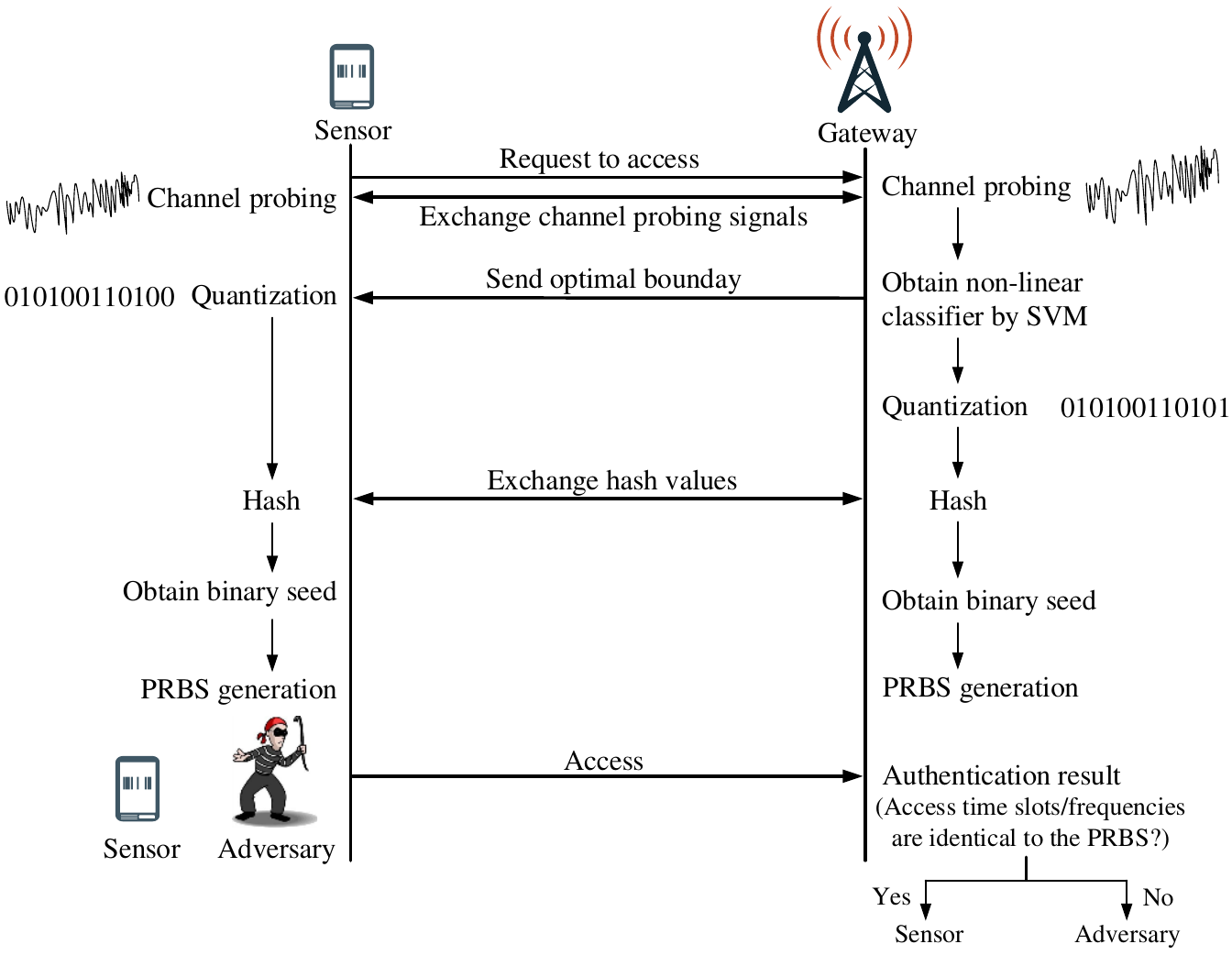}
\caption{Lightweight authentication scheme based on SVM.}
\end{figure}

Fig. 2 characterizes the flow diagram of the developed lightweight authentication scheme.
Firstly, the measurements of selected features are obtained by channel probing.
In order to convert these measurements into binary sequences, we develop a new quantization technique based on the SVM  to  derive an optimal non-linear boundary  to separate the dense data and sparse data. Compared with the received signal strength (RSS)-based quantization technique \cite{11}, our technique reduces wrong decisions through diminishing the measurements near the boundary.
Then the gateway sends the non-linear boundary to the sensor, so that highly similar binary sequences will be acquired on both gateway and
sensor sides because of the channel reciprocity \cite{02}. Hash functions are used for verification, so that identical seed is obtained, and then the same PRBS is generated for authentication between each sensor and the gateway. Explicitly,  the seed generated by the gateway and sensor is concealed from any other devices because of the unique and unpredictable features of communication link used. In this scheme, the AI technique (i.e. SVM) facilitates the security enhancement through deriving a non-linear classifier at the gateway, which equips enough high energy and storage space for training. More importantly, due to the derived optimal non-linear boundary by SVM, the highly similar binary sequences are acquired in the quantization phase. In this way, the seed transmission is not required for verification. Hence, the proposed authentication scheme based on pseudo-random access  is lightweight at sensors in the large-scale IoT network.

\subsection{Holistic Approach based on Trust Management and Online Machine Learning}

In order to achieve the fast authentication and progressive authorization, we envision a trust management based holistic access control scheme by intelligently exploiting the time-varying features of the transmitter, i.e. communication-related, hardware-related attributes and user behaviors,  to improve wireless security. To be more specific, one feature is utilized for initial authentication, thus achieving fast access of the transmitter. To assess the security risks caused by a misdetected adversary, we explore the trust management to establish the security  relationship between the transceiver. Through dividing the IoT services/resources  into $M$ levels,  the transmitter is regulated to access the authorization level corresponding to its trust value,  resulting in a holistic approach.
Then the online machine learning \cite{15} is exploited to  dynamically update the trust relationship relying on real-time detection of the feature estimation at receiver. In achieving this, the decision-making in high layer is also required for modelling the holistic authentication and authorization as well as for improving the security performance.
Through such scheme, the uncertainty  and  cascading risks caused by adversaries  can be evaluated and controlled in real time, so that the progressive protections for IoT devices are achieved. The following algorithm summarizes the proposed holistic approach based on the trust management and online machine learning in the complex dynamic IoT environment.

\begin{figure}[htbp]
\centering
\includegraphics[width=14cm,height=11cm]{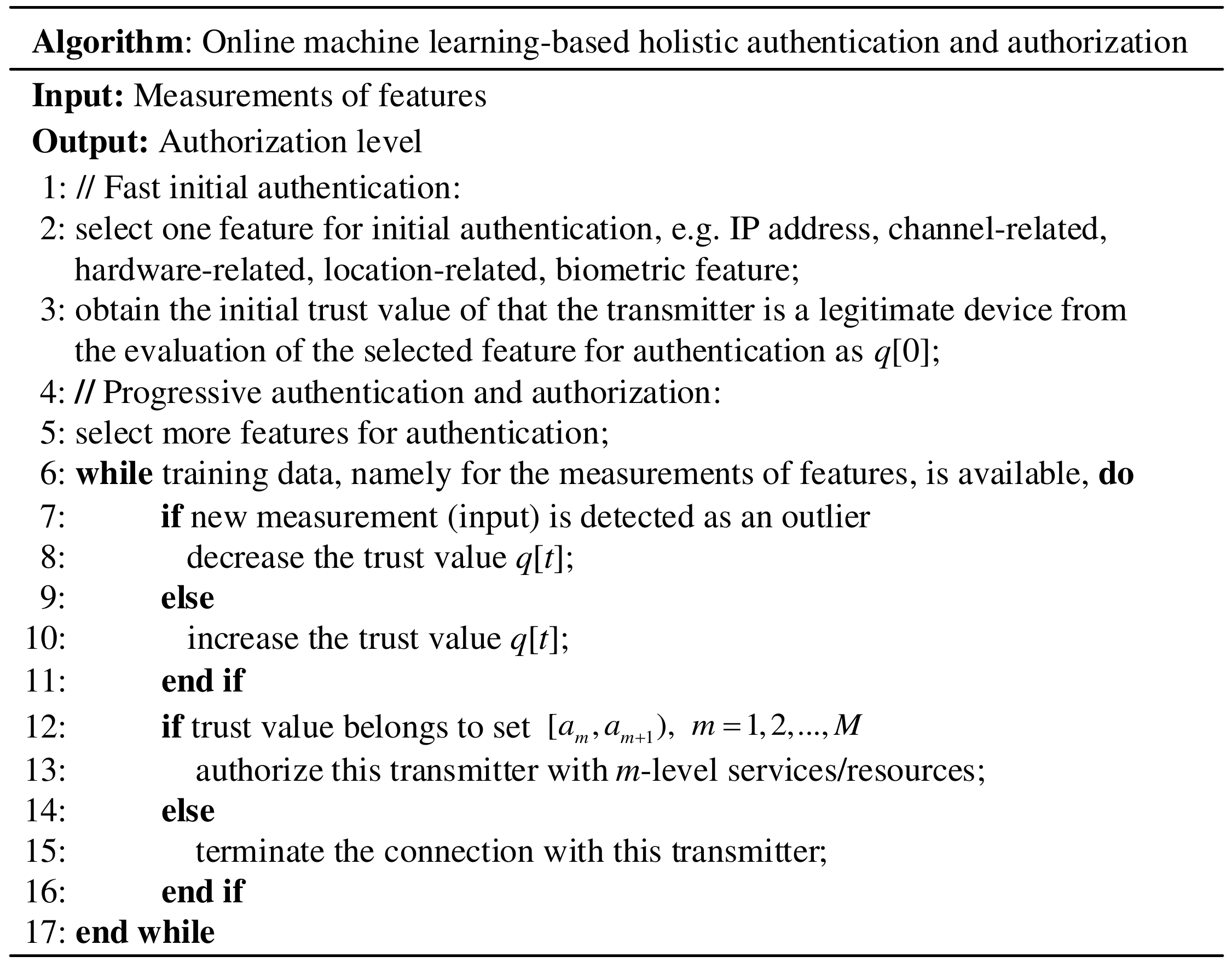}
\end{figure}

Furthermore, the utilization of cooperations among devices \cite{03} may provide seamless protections and enhanced security for large-scale IoT systems. For instance,
the recommendations from other IoT devices may also be exploited for updating the trust relationship between transceiver, so that comprehensive evaluation of the transmitter can be achieved.
The game theory \cite{16}  may be  explored for  studying the interactions among devices to achieve cooperations among devices and to manage trust adaptively.  The basic idea is that the misbehavior and wrong recommendation of a device will be punished. In order to achieve high utility/profit, every device should be more cautious about their behaviors in the large-scale IoT system.  Through seeking the equilibrium, legitimate devices may achieve maximal utility/profit in the presence of adversaries, so that the secure communications in the large-scale IoT system may be  guaranteed. In this scheme, the machine learning techniques and game theory contribute to the security enhancement  by learning the features and behaviors of transmitter, as well as by cooperating with other IoT devices under the limited statistical properties and unpredictable dynamics.

\section{PERFORMANCE ANALYSIS AND EVALUATION}

\begin{figure}[htbp]
\centering
\includegraphics[width=10cm,height=9cm]{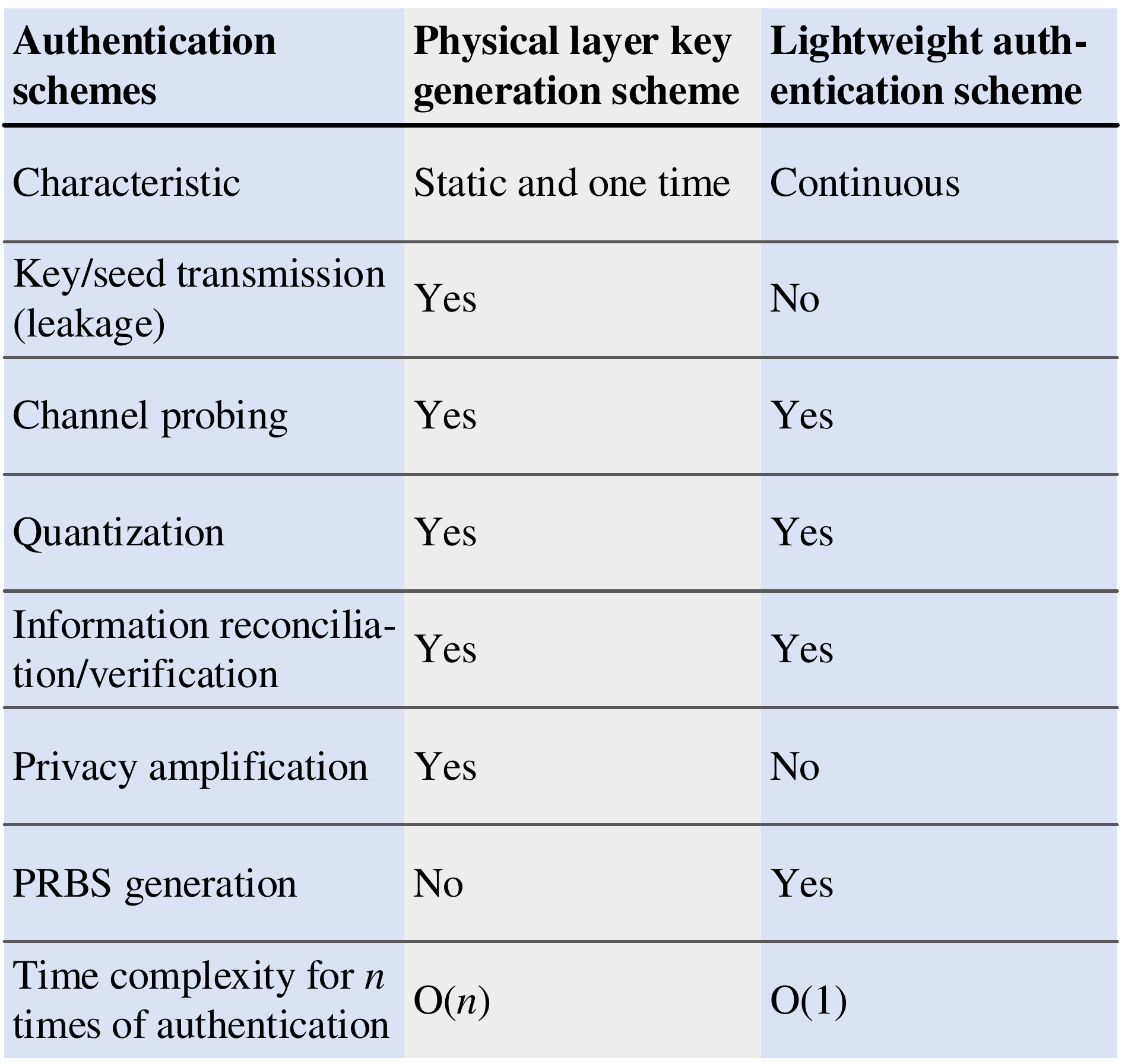}
\caption*{Table I: Comparison results between the physical layer key generation scheme of \cite{02} and the developed lightweight authentication scheme. }
\end{figure}

Table I characterizes the comparison results between the
physical layer key generation scheme of \cite{02} and the developed
lightweight authentication scheme. We can observe
from Table I that the physical layer key generation scheme
of \cite{02} is static in time and one-time authentication method,
while the developed scheme achieves continuous authentication in the
time-domain. Furthermore, the seed transmission is not
required in our scheme, thus providing 100\% privacy protection
of the generated PRBSs. In contrast, key transmission in
the physical layer
key generation schemes leads to the potential key leakage although the privacy amplification is processed. Explicitly, compared with
the physical layer key generation scheme of \cite{02}, the developed authentication
scheme achieves less communications during the key generation/seed acquirement process,  but it needs more computation for generating
PRBS at sensors. More importantly,  the time complexity for $n$  times of authentication in the developed scheme is much lower than that of the physical layer key generation scheme of \cite{02}.  As a conclusion, the developed lightweight authentication scheme achieves
fast authentication and security enhancement in large-scale IoT networks.

\begin{figure}[htbp]
\centering
\includegraphics[width=12cm,height=9.5cm]{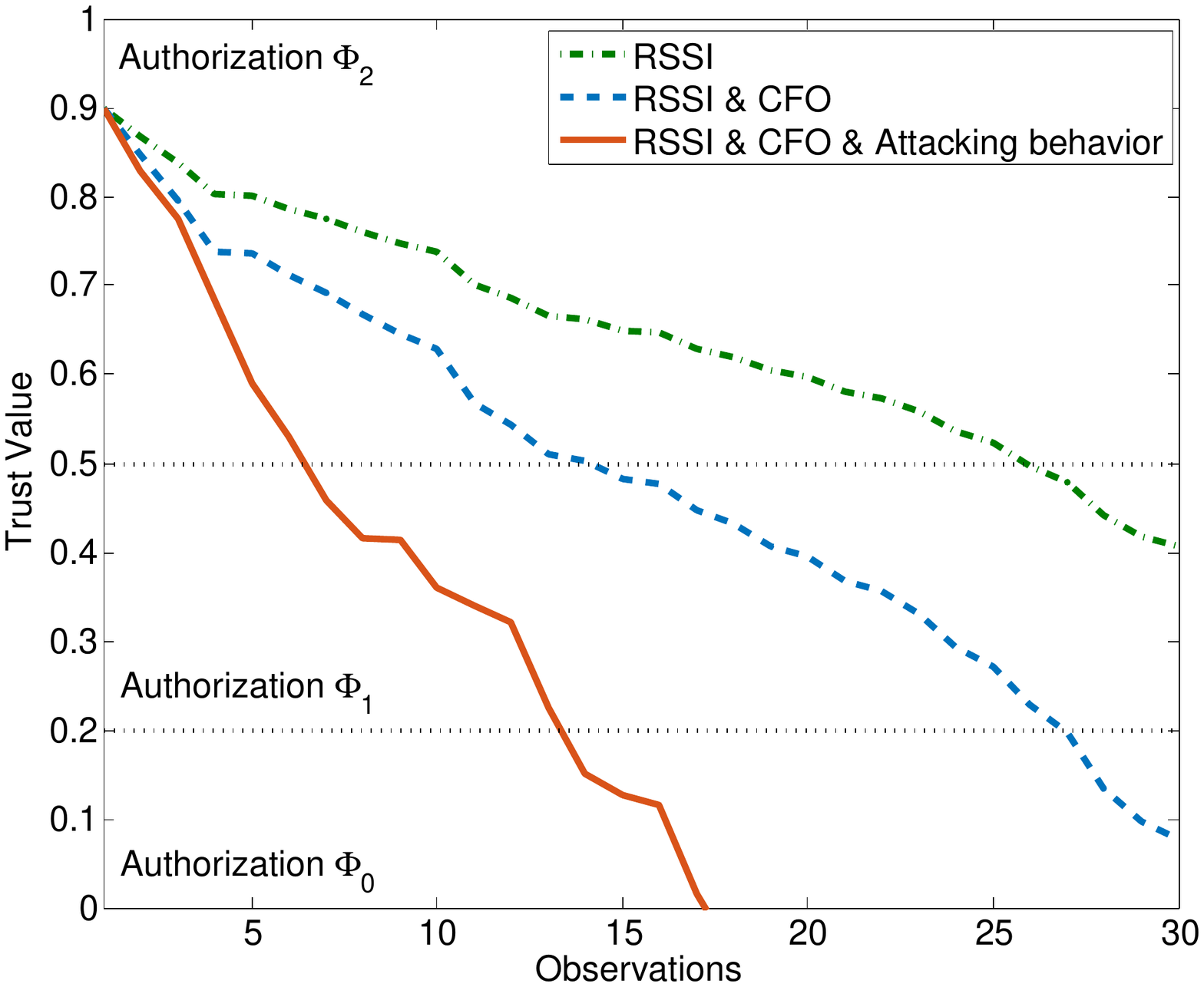}
\caption{Trust value update of a misdetected adversary  in the developed holistic authentication and authorization scheme with respect to the different features observed.
The x-axis is the number of observations of features and attacking behaviors of this adversary obtained by the receiver.}
\end{figure}

In the proposed holistic authentication and authorization scheme, the number of authorization levels is set to 3, where $\Phi_{2}$ represents the set with all services/resources,  set $\Phi_{1}$  contains segmental services/resources, and $\Phi_{0}$ is a null set. Note that the number and thresholds of authorization levels can be designed and adjusted by the users according to the specific IoT application scenarios.
The trust value update of the new authenticated device (namely for a misdetected adversary) is characterized in Fig. 3 with respect to different features: received signal strength indication (RSSI), carrier frequency offset (CFO), and attacking behaviors. Specifically, the attacking behaviors represent the malicious data injection attacks, and are set to be ``on-off" according to the practical communications in large-scale IoT networks. The initial trust value is created based on the evaluation of  the transmitter's IP address. Since the IP address may be faked or forged, it is set to be  0.9.
 We can observe from Fig. 3 that the trust  values of this misdetected adversary decrease faster if more features are considered for the holistic authentication and authorization. The reason for this trend is that the receiver has higher confidence in the authentication result
  when more features and attacking behavior are considered,  thus the punishment   in decreasing the trust
value of this device is designed to be bigger. Once the trust value of this misdetected adversary
is lower than 0.5, it will be authorized to access less services/resources in the large-scale IoT system to reduce the damages and security risks caused by this device as well as to protect the other devices.

\section{CONCLUSION AND FUTURE PERSPECTIVES}
This article firstly introduced the characteristics of large-scale IoT networks, as well as the challenges for the conventional authentication and authorization. The benefits of AI for security enhancement were presented and
some existing AI-based authentication schemes were reviewed.  Then we developed two new AI-enabled  approaches to achieve fast authentication and progressive authorization as well as to enhance security, including the lightweight authentication scheme and holistic access control scheme.
As a conclusion, AI provides a new insight into authentication and authorization in large-scale IoT networks by simplifying the security management, by seeking available information, and
 by adapting to complex dynamic environment for
achieving effective protections for legitimate communications.

With the rapidly increasing number of intelligent devices used in large-scale IoT networks, there are also many other perspectives and areas in which AI can play a remarkable role and improve the quality of human lives. A range of future research ideas on AI for security and IoT services can be summarized as follows.

A distributed IoT system may suffer from false authentications and authorizations of new entities when the trusted nodes are compromised, leading to potential privacy leakage and security risks. To circumvent these impediments, a cooperative access control scheme based on recommendation  may provide security enhancement.
Multiple IoT devices can act as referrers to distributively authenticate a public device and to issue the initial authorization level based on the recommendation results,
even though there is no trusted central party in the IoT system.

Design of more effective machine learning and distributed machine learning algorithms is also beneficial for the IoT applications and security provisioning.
Statistical learning methods ranging from simple calculation of averages to the construction of complex models may be utilized, such as Bayesian learning and maximum-likelihood learning.
Some AI techniques may be also explored for reducing the dimensionality of authentication and authorization systems, such as the kernel machine and principal component analysis.

Considering that the insider attacks caused by adversaries who has passed the authentication may  cause cascaded damages to large-scale IoT systems, the game theory may be also utilized for defending against insider attacks through modelling the behaviors of attackers. Some typical game models for studying interactions between legitimate devices and adversaries include potential game, Stackelberg game, and evolutionary game, just to name a few. The basic idea is that the incentive mechanism and punishment mechanism can be designed for achieving security enhancement and cooperations among untrusted entities in smart IoT applications.

For resource allocation problems in large-scale IoT, we may explore the game theory, auction theory and reinforcement learning to achieve better decision making.
The interactions among devices can be modelled based on game theory, so that cooperations can be consummated  by searching its equilibrium based on learning algorithms. Auction theory, such as ascending-bid auctions and descending-bid auctions, may be utilized for modelling the resource competitions among multiple entities in IoT, so that the reliable ``market" may be built for IoT applications.

\ifCLASSOPTIONcaptionsoff
  \newpage
\fi

\appendices

~\\

\end{document}